\documentstyle[preprint,aps]{revtex}

\begin{document}
\title{Bohr-Sommerfeld Quantization of Periodic Orbits}
\author{G\'abor Vattay\cite{LAbs}}
\address{Division de Physique Th\'eorique, Institut de Physique Nucl\'eaire,\\
F-91406 Orsay Cedex, France\\
e-mail:vattay@ipncls.in2p3.fr}
\date{\today}

\maketitle

\begin{abstract}
We show, that the canonical invariant part of $\hbar$ corrections to the
Gutzwiller trace formula and the Gutzwiller-Voros spectral determinant
can be computed by the Bohr-Sommerfeld quantization rules, which usually
apply for integrable systems.
We argue that the information content of the classical action and stability
can be used more effectively than in the usual treatment.
We demonstrate the improvement of precision on the example of
the three disk scattering system.
\end{abstract}

Gutzwiller trace formula for chaotic systems is often presented
as the counterpart of the Bohr-Sommerfeld (BS) quantization of integrable
systems\cite{Gutzwiller}. Consequently, corrections of the trace formula
proportional with
powers of $\hbar$ are usually associated with quantum
corrections\cite{Gaspard}.
In this letter we would like to show, that a part of $\hbar$
corrections is not connected to deep quantum effects and they
can be calculated with some precision from purely semiclassical BS
arguments. Then we propose a new trace formula and spectral determinant
which is more precise than the usual trace formula although
it uses only the linear stability and action as an input data,
just like the original trace formula.

First we would like to introduce periodic orbits from an
unusual point of view. Chaotic and integrable systems on the level
of periodic orbits are in fact not
as different from each other as we might think. If we start orbits in the
neighborhood of a periodic orbit and look at the picture on the Poincar\'e
section we can see a regular pattern.  For stable periodic orbits the points
form small ellipses around the center and for unstable orbits they form
hyperbola. The motion close to a periodic orbits is
regular in both cases. This is due to the fact, that we can linearize the
Hamiltonian close to a periodic orbit, and linear systems are always
integrable. Based on Poincar\'e's idea, Arnold and coworkers have
shown\cite{Arnold}, that the
Hamiltonian close to a periodic orbit can be brought to a very practical
form.  One has to introduce new coordinates: one which is parallel with the
orbit
($x_{\parallel}$) and others which are orthogonal. In the orthogonal
directions we get linear equations. These equations with $x_{\parallel}$
dependent rescaling can be transformed into normal coordinates so that we get
tiny oscillators, or inverse oscillators, in the new coordinates with constant,
frequencies. In the new coordinates, the Hamiltonian is
\begin{equation}
H_0(x_{\parallel},p_{\parallel},x_n,p_n)=\frac{1}{2}
p_{\parallel}^2+U(x_{\parallel})+\sum_{n=1}^{d-1}
\frac{1}{2}(p_n^2\pm\omega_n^2 x_n^2),\label{ArPo}
\end{equation}
which is one possible normal form of the Hamiltonian in the neighborhood
of a periodic orbit.
The $\pm$ sign denotes, that for stable modes the oscillator potential is
positive, while for an unstable mode it is negative. Since the eigenvalues
of the monodromy or Jacobi stability matrix of a periodic orbit are invariant
under the transformations we made sofar, the oscillator frequencies can be
expressed for unstable modes with the  Ljapunov exponent of the orbit
\begin{equation}
\omega_n=\ln|\Lambda_{p,n}|/T_p,
\end{equation}
where $\Lambda_{p,n}$ is the expanding eigenvalue of the Jacobi
matrix and $T_p$ is the period of the orbit.
Also, for stable directions the eigenvalues of the Jacobi matrix are
connected with $\omega$ as
\begin{equation}
\Lambda_{p,n}=e^{-i\omega_n T_p}.
\end{equation}

The Hamiltonian (\ref{ArPo}) is integrable and can be semiclassically
quantized by the BS rules. The result of the BS
quantization for the oscillators gives the energy spectra
\begin{eqnarray}
E_n&=&\hbar\omega_n\left(j_n+\frac{1}{2}\right)\;\;\;\; \mbox{for stable
modes,}\\
\nonumber
E_n&=&-i\hbar\omega_n\left(j_n+\frac{1}{2}\right) \;\;\;\; \mbox{for unstable
modes,}
\label{har}
\end{eqnarray}
where $j_n=0,1,...$.
It is convenient to introduce the index $s_n=1$ for stable and
$s_n=-i$ for unstable directions.
The parallel mode can be quantized implicitly trough the classical
action function :
\begin{equation}
\frac{1}{2\pi}\oint p_{\parallel}dx_{\parallel}=
\frac{1}{2\pi}S_{\parallel}(E_m)=\hbar\left(m+\frac{\mu_p\pi}{2}\right),
\label{mil1}
\end{equation}
where $\mu_p$ is the Maslov index of the motion in the parallel
direction.
This latter condition can be rewritten in
the equivalent form
\begin{equation}
(1-e^{iS_{\parallel}(E_m)/\hbar-i\mu_p\pi/2})=0.\label{Bohr}
\end{equation}
The eigenenergies of a semiclassically quantized periodic
orbit\cite{Miller}
are all the possible energies
\begin{equation}
E=E_m+\sum_{n=1}^{d-1} E_n.
\label{mil2}
\end{equation}
This relation allows us to change in (\ref{Bohr}) $E_m$ with the full energy
minus the oscillator energies $E_m=E-\sum_n E_n$.
All the possible
eigenenergies of the periodic orbit then are the zeroes of the expression
\begin{equation}
\Delta_p^{BSH}(E)=\prod_{j_1,...,j_{d-1}}(1-e^{iS_{\parallel}
(E-\sum_n\hbar s_n\omega_n(j_n+1/2))/\hbar
-i\mu_p\pi/2}).\label{spede}
\end{equation}

Now, we show how one can derive the Gutzwiller trace formula from
(\ref{spede}).
We have to expand the action around $E$ to first order
$S_{\parallel}(E+\epsilon)\approx S_p(E)+T_p(E)\epsilon$,
where $T_p(E)$ and $S_p(E)$ are the period and the action of the orbit, and
we have to use the relations
of $\omega$-s and the eigenvalues of the Jacobi matrix,
we get
\begin{equation}
\Delta_p(E)=\prod_{j_1,...,j_{d-1}}\left(1-
\frac{e^{iS_p(E)/\hbar -i\nu_p\pi/2}}{\prod_n|\Lambda_{p,n}|^{1/2}
\Lambda_{p,n}^{j_n}}
\right)\label{GV},
\end{equation}
where $\nu_p$ is the Maslov index of the orbit.
Now, if we have many primitiv orbits and we would like to
construct a function formally, which is zero whenever the energy coincides
with the BS quantized energy of one of the periodic orbits, we
have to take the product of these determinants:
\begin{equation}
\Delta(E)=\prod_p\Delta_p(E).
\end{equation}
This is exactly the Gutzwiller-Voros\cite{Voros,Selberg} spectral determinant,
which is the regularized semiclassical expression for the spectral determinant
$$\Delta(E)=\det(E-\hat{H})$$ of the Hamilton operator of the system.
The logarithmic derivative of this quantity gives the trace of the
Green's function and the oscillating part of the trace formula
in semiclassical approximation
\begin{equation}
Tr G(q',q',E)=-\frac{d}{dE}\log\Delta(E)\approx\frac{1}{i\hbar}
\sum_{p,r}T_p(E)\frac{e^{irS_p(E)-ir\nu_p\pi/2}}{|\det(1-J_p^r)|^{1/2}},
\end{equation}
where the summation goes for the primitive periodic orbits and their
repetitions, $J_p$ is the monodromy or Jacobi matrix of the periodic
orbit. The construction of the spectral determinant (\ref{spede})
described above is not unique in the sense, that we can multiply each
periodic orbit's spectral determinant with a smooth function which
does not have zeroes and poles on the complex energy plane.
However, the logarithmic derivative of such a smooth function will not
contribute to the oscillating part of the trace formula and can be
considered as a part of its smooth part.

{}From this derivation we can see, that the Gutzwiller trace formula
is recovered only if we linearize the action.
This is a very bad approximation for low energies, where the
ratio $T_p(E)/S_p(E)\sim 1/E,$
is large. In that regime we might expect that the spectral
determinant (\ref{spede}) works better than the Gutzwiller-Voros
formula (\ref{GV}).

That this is the case, we have checked on the example of the three disk
scattering  system\cite{Rice,CE}
at the standard parameters of disk separation 6 compared to the radius of
the disks. This is a billiard system, where the parallel action is
$S(E)=kL_p,$ where $k=\sqrt{E}$ is the wave number
(with mass unit $m=1/2$ ) and $L_p$ is the geometrical length of the the orbit.
There is only one unstable oscillator mode with oscillator energies
$$E_{p,j}(k)=-i\hbar\frac{2k\log|\Lambda_p|}{L_p}(j+1/2).$$
The new spectral determinant as a function of the wave number is
\begin{equation}
\Delta^{BSH}(k)=\prod_p\prod_j(1-e^{iL_p\sqrt{k^2-E_{p,j}(k)}/\hbar-i\nu_p\pi/2}).
\label{uj}
\end{equation}
If we expand the exponent in $\hbar$ we can see that it gives
corrections to the leading action and stability term. We can
compare the first $\hbar$ correction with the exact $\hbar$ correction
computed in Ref.\cite{Gaspard} and Ref.\cite{VR}. The $\hbar$ corrected
spectral determinant with the exact correction is defined as
\begin{equation}
\Delta^{ex}(k)=\prod_p\prod_j(1-e^{ikL_p/\hbar -i\nu_p\pi/2 -(j+1/2)\ln
|\Lambda_p| +
i\hbar C^{ex (1)}_{p,j}/k+...}),
\end{equation}
where $C^{ex (1)}_{p,j}$ the exact first $\hbar$ correction.
The first correction coming from (\ref{uj}) is
$$C^{BSH (1)}_{p,j}=(j+1/2)^2(\ln |\Lambda_p|)^2/L_p.$$
On Fig. 1 we can see, that this accounts for about $80\%$ of the exact
correction. As we will see later, this is not the whole first
$\hbar$ correction which comes form the BS quantization,
it illustrates only that by using (\ref{uj}) we already take into account
a whole series of $\hbar$ corrections.
We also compared the exact quantum mechanical resonances\cite{Wirzba} with
those computed\cite{HHR} from the Gutzwiller-Voros spectral determinant
and with the Gutzwiller-Voros spectral determinant
with the first $\hbar$ correction\cite{VR} and plotted the
results on the complex wave number plane. We can see
on Fig. 2,
that for large $k$ the Gutzwiller-Voros spectral determinant, its
$\hbar$ corrected version and the new BS type expressions
approximate the resonances accurately, with a few percent error. However,
the lowest resonances are approximated better by the new expression and
the $\hbar$ corrected Gutzwiller-Voros determinant is even worse than the
uncorrected.
This is because the Gutzwiller-Voros determinant and its corrected
version are asymptotic series expanded in the powers
of $1/k$, while the new formula approximates the eigenenergies of
the individual periodic orbits for small values of $k$ also correctly.
We can conclude, that by using (\ref{uj}) or in general (\ref{spede})
we can considerably improve the Gutzwiller-Voros theory for low
energies by using exactly the same input data (stability, action,
Maslov index) in a more economic way.

The semiclassical analysis outlined above can be done in a more general
framework by the systematic application of the normal form theory of Birkhoff
and Gustavson. This can be considered as a generalization of the normal
form quantization of Swimm, Delos and Robnik\cite{SDR} from equilibrium
points to the periodic orbits.
The full Hamiltonian expanded close to a
periodic orbit can be written as the
`harmonic' plus an `anharmonic' perturbation
\begin{equation}
H(x_{\parallel},p_{\parallel},x_n,p_n)=H_0(x_{\parallel},p_{\parallel},x_n,p_n)
+H_A(x_{\parallel},x_n,p_n),
\end{equation}
where the anharmonic part can be written as a sum of homogeneous
polynomials of $x_n$ and
$p_n$ with $x_{\parallel}$ dependent coefficients:
\begin{eqnarray}
H_A(x_{\parallel},x_n,p_n)&=&\sum_{k=3}^{\infty}
H^{k}(x_{\parallel},x_n,p_n),\\
H^{k}(x_{\parallel},x_n,p_n)&=&\sum_{\sum_n l_n+m_n=k}
h^{k}_{l_n,m_n}(x_{\parallel})x_n^{l_n}p_n^{m_n}\label{ArPoAh}.
\end{eqnarray}
We can carry out the canonical transformations introduced by Birkhoff and
Gustavson\cite{BG,SDR,Gutzwiller} and the Hamiltonian can be brought to
normal form in the orthogonal directions. The $x_{\parallel}$ coordinate
plays the role of a parameter. After the transformation up to
order $N$ the Hamiltonian (\ref{ArPoAh}) is
\begin{equation}
H(x_{\parallel},p_{\parallel},\tau_1,...\tau_{d-1})=
H_0(x_{\parallel},p_{\parallel},\tau_1,...,\tau_{d-1})+
\sum_{j=2}^{N} U^{j}(x_{\parallel},\tau_1,...,\tau_{d-1})+h.o.t,
\end{equation}
where $U^{j}$ is a $j$-th order homogeneous polynomial of $\tau$-s with
$x_{\parallel}$ dependent coefficients and
$\tau_n=\frac{1}{2}(p_n^2\pm\omega_n^2 x_n^2)$
is the Hamiltonian function of the original oscillator. This Hamiltonian
truncated at order $N$ is integrable, the nonintegrability is pushed
to the higher order terms (h.o.t) .
The orthogonal part can then be BS quantized by quantizing the
individual oscillators, replacing $\tau$-s as we did in (\ref{har}). This
leads to a one dimensional
effective potential indexed by the {\em quantum numbers} $j_1,...,j_{d-1}$
\begin{eqnarray}
&&H(x_{\parallel},p_{\parallel},j_1,...,j_{d-1})=
\frac{1}{2}p_{\parallel}^2+U(x_{\parallel})+\sum_{n=1}^{d-1}
\hbar s_n \omega_n (j_n+1/2)+\\ \nonumber
&&+\sum_{k=2}^N U^{k}(x_{\parallel}, \hbar s_1 \omega_1(j_1+1/2),
\hbar s_2 \omega_2 (j_2+1/2),...,
\hbar s_{d-1}\omega_{d-1}(j_{d-1}+1/2)),\label{energy}
\end{eqnarray}
where $j_n$ can be any non-negative integer. The term with index
$k$ is proportional with $\hbar^k$ due to the homogeneity of the polynomials.
The parallel mode now can be BS quantized for any given set of
$j$-s
\begin{eqnarray}
&&S_p(E,j_1,...,j_{d-1})=\oint p_{\parallel}dx_{\parallel}=\\ \nonumber
&&=\oint dx_{\parallel}\sqrt{2\left(E-\sum_{n=1}^{d-1}\hbar
s_n\omega_n(j_n+1/2)-
U_{eff}(x_{\parallel},j_1,...,j_{d-1})\right)}=2\pi \hbar
\left(m+\frac{\mu_p\pi}{2}
\right),
\end{eqnarray}
where $U_{eff}$ contains all the $x_{\parallel}$ dependent terms of the
Hamiltonian. The spectral determinant becomes
\begin{equation}
\Delta^{BSA}(E)=\prod_p\prod_{j_1,...,j_{d-1}}
(1-e^{iS_p(E,j_1,...,j_{d-1})/\hbar-i\nu_p\pi/2}).
\end{equation}
Here one can see, that the indices $j$, which were just auxiliary indices
in in the Gutzwiller-Voros approach, now can be interpreted as "orthogonal
quantum numbers". Formally, the integrability of the Hamiltonian is maintained
in each level of the approximation and the Ruelle\cite{Ruelle} type zeta
functions $$\zeta^{-1}_{j_1,j_2,...,j_{d-1}}(E)=\prod_p
(1-e^{iS_p(E,j_1,...,j_{d-1})/\hbar-i\nu\pi/2}),$$
corresponding to a given $j$ configuration,
can be interpreted as the factorization of the total spectral determinant
according to subspaces with fixed quantum numbers.
If we expand $S_p$ in the exponent in the powers of $\hbar$
$S_p=\sum_{k=0}^{N} \hbar^k S_k,$ we
get
corrections to the Gutzwiller-Voros spectral determinant in all powers
of $\hbar$.
There is a very attracting feature of this semiclassical expansion.
$\hbar$ in $S_p$ shows up only in the combination $\hbar s_n \omega_n (j_n
+1/2)$.
A term proportional with $\hbar^k$ can only be a homogeneous
expression of the oscillator energies $s_n\omega_n (j_n+1/2)$.
We mention here Ref.\cite{Niall} as a good example, where the
superiority of the method outlined here can be demonstrated
above the pure periodic orbit theory and the comparison
of $\hbar$ expansion and BS quantization can be
clearly studied.

The $\hbar$ corrections derived  here are {\em doubly} semiclassical,
since they give semiclassical corrections to the semiclassical
approximation. What can quantum mechanics add to this ? Since quantum
mechanics is not invariant for canonical transformations, the
derived $\hbar$ corrections give only the leading behaviour of
corrections and the exact corrections can be computed by other
methods. The Birkhoff-Gustavson transformations should be
replaced by quantum perturbation theory and semiclassical quantum numbers
should be replaced by exact quantum numbers. This has been done in
Ref.\cite{VR} which we are going to publish elsewhere. We don't think,
that the semiclassical determination of $\hbar$ corrections in higher
orders is a very practical way to compute them, but in predicting
the general structure of the corrections it helps us to
understand their general behaviour and later probably to sum them up in
order to get meaningful formulas also for low energies.

The author thanks the discussions with E. Bogomolny, O. Bohigas,
P. Cvitanovi\'c, P. E. Rosenqvist, N. Whelan and A. Wirzba. This project
has been financed by the EHCM PECO, OTKA F17166 and T17493.

\begin{figure}
\caption{
Comparison between the exact $\hbar$ corrections and the
BS approximation. We have taken all the periodic
orbits $p$ of the three disk system up to 9 bounces. We computed
the first $\hbar$ correction for $j=0$ ($C^{ex}_{p,0})$. Then we computed the
quantity $(\ln|\Lambda_p|)^2/4L_p$, which correction comes from
the lowest order harmonic approximation of the BS
quantization of the periodic orbit ($C^{BSH (1)}_{p,0}$). We plotted here this
value versus the exact $\hbar$ correction for each periodic
orbit $p$. We can see, that for the 22 shortest orbits
they are almost linearly correlated and about $80\%$ of the
exact correction comes from this effect. The rest is coming from
the anharmonicity and from deep quantum effects.}
\end{figure}

\begin{figure}
\caption{Complex resonances of the 3 disk scattering system.
The quantum calculation, the Gutzwiller-Voros approximation,
the Gutzwiller-Voros approximation with one $\hbar$ correction
added and the BS spectral determinant in harmonic
approximation (BSH) . We used all the periodic orbits up to 9 bounces.
For small $Re k$ resonances the $\hbar$ correction breaks down,
while our approximation is deviating the right way from the
Gutzwiller-Voros result.  }
\end{figure}

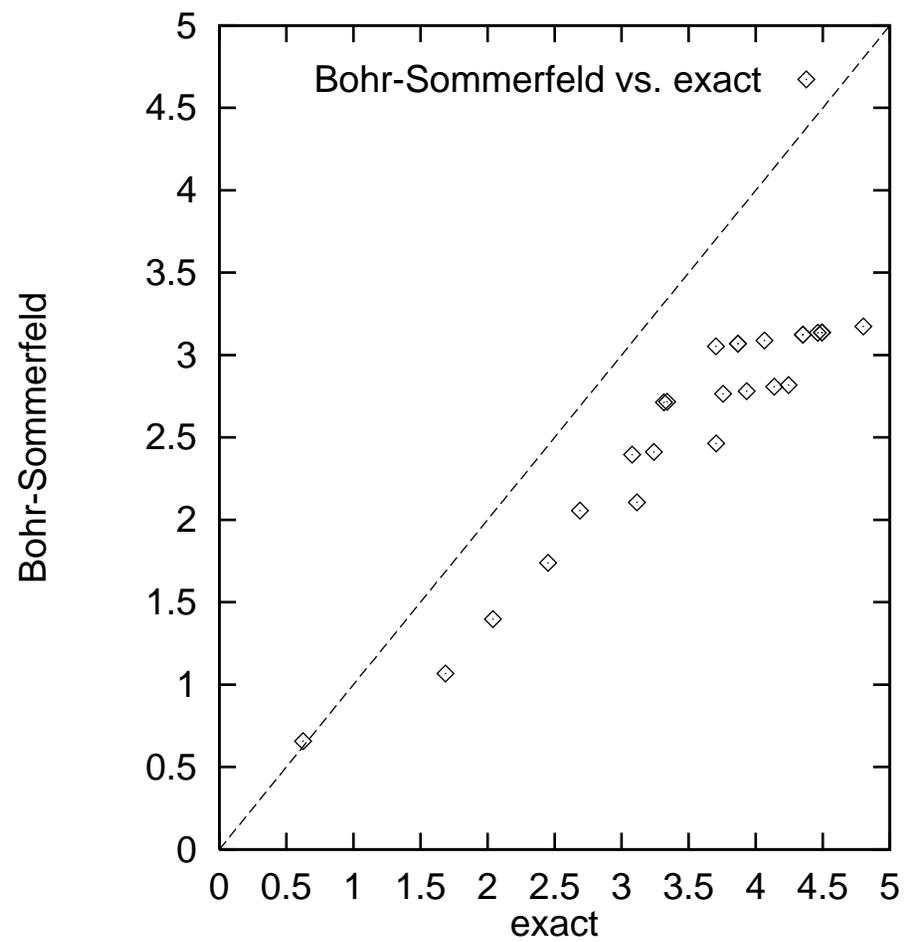
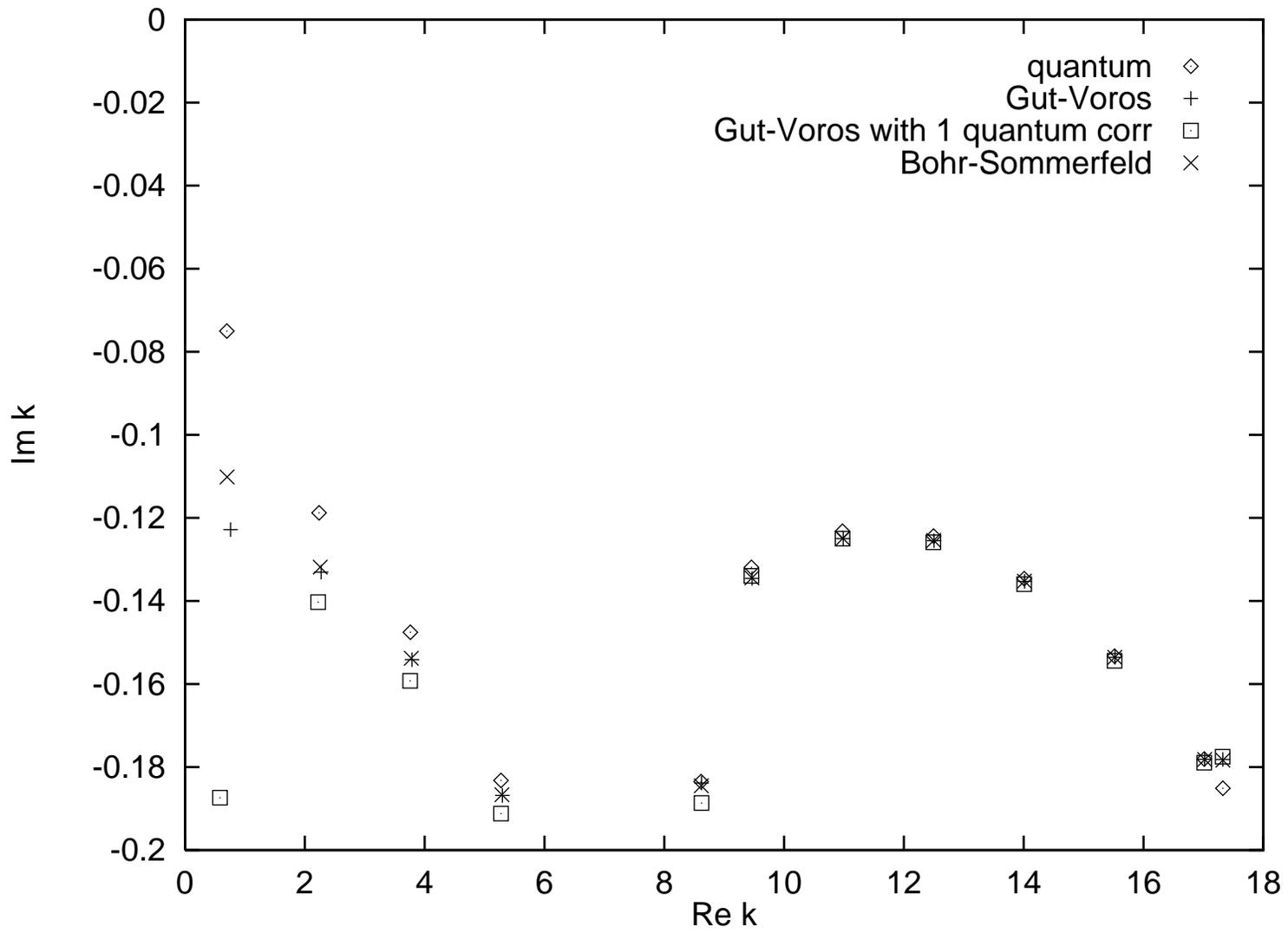
\end{document}